\documentclass[aps,twocolumn,showpacs,preprintnumbers,floatfix,superscriptaddress]{revtex4}
\usepackage{times}
\usepackage{mathptmx}
\usepackage{bm}
\usepackage{graphicx}
\usepackage{dcolumn}
\usepackage{tabularx}

\begin{document}

\preprint{APS/123-QED}
\title{Pseudospin Soliton in the $\nu=1$ Bilayer Quantum Hall State
}
\author{A. Fukuda}
\affiliation{Research Center for Low Temperature and Materials Sciences, Kyoto
University, Kyoto 606-8502, Japan}
\author{D. Terasawa}
\affiliation{Graduate School of Science, Department of Physics, Tohoku University, Sendai
980-8578, Japan }
\author{M. Morino}
\affiliation{Graduate School of Science, Department of Physics, Tohoku University, Sendai
980-8578, Japan }
\author{K. Iwata}
\affiliation{Graduate School of Science, Department of Physics, Kyoto University, Kyoto
606-8502, Japan }
\author{S. Kozumi}
\affiliation{Graduate School of Science, Department of Physics, Tohoku University, Sendai
980-8578, Japan }
\author{N. Kumada}
\affiliation{NTT Basic Research Laboratories, NTT Corporation, 3-1 Morinosato-Wakamiya,
Atsugi 243-0198, Japan}
\author{Y. Hirayama}
\affiliation{Graduate School of Science, Department of Physics, Tohoku University, Sendai
980-8578, Japan }
\affiliation{NTT Basic Research Laboratories, NTT Corporation, 3-1 Morinosato-Wakamiya, Atsugi 243-0198, Japan}
%%\affiliation{SORST-JST, 4-1-8 Honmachi, Kawaguchi, Saitama 331-0012, Japan}
\author{Z. F. Ezawa}
\affiliation{Graduate School of Science, Department of Physics, Tohoku University, Sendai
980-8578, Japan }
\author{A. Sawada}
\affiliation{Research Center for Low Temperature and Materials Sciences, Kyoto
University, Kyoto 606-8502, Japan}

\date{\today}

\begin{abstract}
We investigate a domain structure of pseudospins, a soliton lattice in the bilayer quantum Hall state at total Landau level filling factor $\nu =1$, in a tilted magnetic field, where the pseudospin represents the layer degree of freedom. An anomalous peak in the magnetoresistance $R_{xx}$ appears at the transition point between the commensurate and incommensurate phases. The $R_{xx}$ at the peak is highly anisotropic for the angle between the in-plain magnetic field $B_\parallel $ and the current, and indicates a formation of the soliton lattice aligned parallel to $B_\parallel $.
Temperature dependence of the $R_{xx}$ peak reveals that the dissipation is caused by thermal fluctuations of pseudospin solitons.
We construct a phase diagram of the bilayer $\nu =1$ system as a function of $B_\parallel$ and the total electron density. We also study effects of density imbalance between the two layers.
\end{abstract}

\pacs{73.43.-f,73.43.Nq,73.43.Qt}
\maketitle

%\email{fukuda@scphys.kyoto-u.ac.jp}




Two-dimensional electron gas (2DEG) in a strong magnetic field is an ideal system to investigate many-body phenomena. Since the quantization of electron motion into Landau levels (LLs) quenches the kinetic energy, electron-electron interactions dominate the physics. When two or more LLs are brought close in energy near the Fermi level, the Coulomb interaction leads to a broken-symmetry state, which can be described as a new class of ferromagnet \cite{DasSarma_book,Ezawa_book}. This is best illustrated in the bilayer quantum Hall (QH) state at total LL filling factor $\nu =1$. When two 2DEGs are set close, even in the limit of zero tunneling energy, strong interlayer interactions produce a broken-symmetry state with spontaneous interlayer phase coherence \cite{DasSarma_book,Ezawa_book}.
A number of interesting phenomena, such as Josephson-like interlayer tunneling \cite{Spielman} and vanishing Hall resistance for counterflowing currents in the two layers \cite {Kellogg_counterflow,Tutuc_counterflow}, have been observed.
This state can be viewed in several ways, including Bose condensate of interlayer excitons\thinspace \cite{Eisenstein} and pseudospin ferromagnet \cite {DasSarma_book,Ezawa_book,Yang}, where the pseudospin represents the layer degree of freedom, in which the pseudospin up and down denotes electrons in the front and back layers, respectively. 

In-plane magnetic field $B_\parallel$ has been used to change pseudospin properties. Murphy {\it et al.} showed evidence of a $B_\parallel $-induced phase transition in the bilayer $\nu =1$ QH state: as $B_\parallel $ is increased, the activation energy gap drops when $B_\parallel $ is smaller than a critical value $B_\parallel ^{\rm C}$ and then stays almost at a constant value for larger $B_\parallel $\cite{Murphy}.
This phase transition is understood as a commensurate-incommensurate (C-IC) transition \cite{Yang}.
The presence of $B_\parallel$ periodically shifts the interlayer phase difference  of electrons between the two layers $\varphi$.
In pseudospin language, while the $z$-component of pseudospin, $P_{z}$, vanishes at the balanced density configuration, the in-plain components of the pseudospin, $P_{x}$ and $P_{y}$, are related to the interlayer phase difference: $\varphi=\arctan (P_{y}/P_{x})$.
For $B_\parallel <B_\parallel^{\rm C}$, pseudospins rotate along the planar direction following the periodically shifting $\varphi$. This is the C phase, where the tunneling energy is minimized.
In the C phase, since neighboring pseudospins are no longer parallel, the pseudospin exchange energy increases with $B_\parallel $. 
In the limits of large $B_\parallel $, pseudospins are uniformly polarized to minimize the exchange energy.
This is the IC phase in the large $B_\parallel $ limit.
Pseudospin configurations in these phases are illustrated in Fig.\thinspace\ref{PD}.
Theories have suggested that in the IC phase for any finite $B_\parallel >B_\parallel ^{\rm C}$ there exists a domain structure of pseudospins\cite{Read,Cote_SL,Yang2,Brey,Ezawa_b,Hanna,ParkPRB} (also illustrated in Fig.\thinspace\ref{PD}). A domain structure is theoretically derived from the sine-Gordon equation, as well as from an isolated soliton solution. Thus we call the domain structure a "soliton lattice" and the domain wall a "pseudospin soliton".
At the pseudospin soliton, a direction of pseudospin $\varphi$ slip by $2\pi$ around a magnetic flux penetrating between the two layers.
Therefore $\varphi$ has a repetitive stepwise function of the position.
An abrupt change of $\varphi$ over the small distance at the domain wall costs the large gradient energy and causes repulsive interactions between solitons, which stabilize the pseudospin soliton into a soliton lattice in the low temperature limit.
The formation and properties of the soliton lattice are directly related to the pseudospin ferromagnetism, and their experimental investigation is essential for understanding the bilayer $\nu =1$ QH state.

In this Letter, we report the observation of an anomalous peak in the longitudinal resistance $R_{xx}$ with a well-developed QH plateau in the Hall resistance $R_{xy}$ around the C-IC transition point in the bilayer $\nu =1$ QH state.
The $R_{xx}$ at the peak changes with the angle $\phi $ between the direction of $B_\parallel $ and the current $I$, following a sinusoidal function.
We interpret this anisotropic transport as a formation of the soliton lattice aligned parallel to $B_\parallel $.
Temperature dependence of the $R_{xx}$ peak reveals that the dissipation is caused by thermally fluctuating pseudospin solitons.
We construct a phase diagram of the bilayer $\nu =1$ QH system as a function of $B_\parallel$ and the total electron density $n_{\rm T}$.
We also show that pseudospin solitons disappear when the bilayer system is off-balanced.

\begin{figure}[t]
\includegraphics[width=\linewidth]{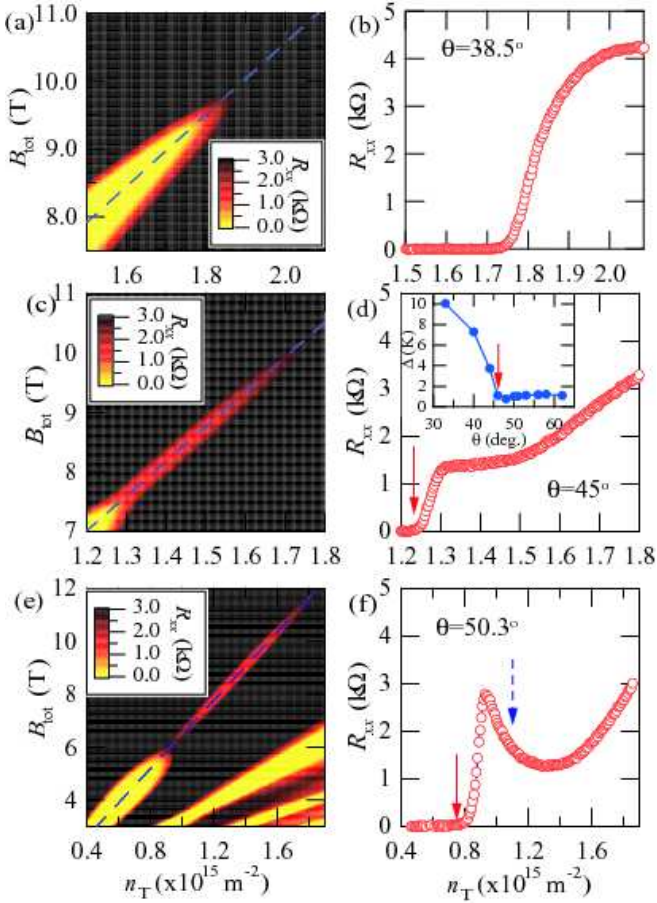}
\caption{Color-scale plots of $R_{xx}$ as a function of the total magnetic field $B_{\rm tot}$ and the total electron density $n_{\rm T}$ at (a) $\theta=38.5^{\rm o}$, (c) $45^{\rm o}$, and (e) $50.3^{\rm o}$ for sample A. 
$B_\parallel $ is applied perpendicular to $I$ ($\phi =90^\circ $).
Data was obtained by simultaneously sweeping the front and back-gate biases to keep the double-quantum-well potential symmetric.
The $R_{xx}$ just at $\nu=1$ along dashed lines is plotted in (b) for $\theta=38.5^{\rm o}$, (d) for $45^{\rm o}$ and (f) for $50.3^{\rm o}$. Temperature is 130 mK for (a)-(d) and 200 mK for (e) and (f). Inset in (d): Activation energy $\Delta$ as a function of $\theta$ at $n_{\rm T}=1.24 \times 10^{15}$\,m$^{-2}$ for sample A. }
\label{DM}
\end{figure}

\begin{figure}[t]
\includegraphics[width=\linewidth]{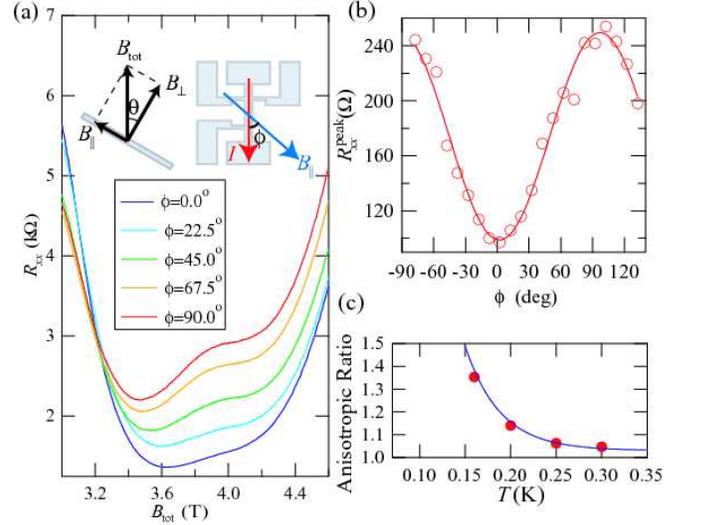}
\caption{(a) $R_{xx}$ as a function of $B_{\rm tot}$ for several values of $\phi$ for sample B. Tilting angle is $\theta = 54^{\rm o}$, the total density is $n_{\rm T} = 0.58 \times 10^{15}$\,${\rm m}^{-2}$, and temperature is $T = 300$ mK.
Inset: Illustrations of the tilting angle $\theta$ and the azimuthal angle $\phi$. (b) $R_{xx}^{\rm peak}$ as a function of $\phi$ at $\theta = 53.5^{\rm o}$, $n_{\rm T} = 0.58\times 10^{15}$\,m$^{-2}$, $B_{\rm tot} = 3.92$\,T, and $T = 190$\,mK.
Solid line is a fitted sinusoidal curve.
(c) Temperature dependence of anisotropic ratio $[R_{xx}(\phi=90^{\rm o}) - R_{xx}(\phi=0^{\rm o})]/ R_{xx}(\phi=0^{\rm o})$ at $\theta = 54^{\rm o}$, $n_{\rm T} = 0.50\times 10^{15}$m$^{-2}$, and $B_{\rm tot} = 3.70$\,T.
Solid line is a guide for eyes.}
\label{Anisotropy}
\end{figure}

We used two double-quantum-well samples with 20-nm-wide GaAs quantum wells.
The tunneling energy is $\Delta_{\rm SAS}=11$ K for sample A and 8 K for sample B\cite{noteDSAS}. 
Low-temperature mobility is $1.0\times 10^{2}$\,m$^2$/Vs at $n_{\rm T}=1.0 \times 10^{15}$\,m$^{-2}$ for both samples. 
By adjusting the front- and back-gate biases, we can independently control $n_{\mathrm{T}}$
and the density imbalance $\sigma \equiv (n_{\mathrm{f}}-n_{\mathrm{b}})/n_{\mathrm{T}}$ between the two layers, where $n_{\mathrm{f}}$ ($n_{\mathrm{b}}$) denotes the electron density in the front (back) layer.
The samples were mounted in the mixing chamber of a dilution refrigerator with a base temperature of 40\thinspace mK.
Measurements were performed using standard low-frequency AC lock-in techniques.
To apply $B_{\parallel }$, the samples were tilted in the magnetic field $B_{\mathrm{tot}}$ by a goniometer with a superconducting stepper motor\thinspace \cite{Suzuki}.
We define the tilting angle $\theta $ as $\tan \theta =B_{\parallel}/B_{\perp }$, where $B_{\perp }$ is the perpendicular magnetic field.
To investigate anisotropic transport, we placed sample B on a two-axis goniometer, for which both $\theta$ and $\phi$ can be controlled independently (the angles are given in the inset of Fig.\thinspace\ref{Anisotropy}(a)).

In Figs.\thinspace\ref{DM}(a), (c), and (e), we present $R_{xx}$ as a function of $B_{\mathrm{tot}}$ and $n_{\rm T}$ near the $\nu=1$ QH state for three tilted angles in a balanced density condition ($\sigma =0$).
Yellow areas indicate small $R_{xx}$ and thus QH states. Dashed lines lie just on the filling factor $\nu=1$.
The $R_{xx}$ at $\nu =1$ (along the dashed lines) is plotted as a function of $n_{\rm T}$ in Figs.\thinspace\ref{DM}(b), (d), and (f).
When $\theta$ is small (Figs.\thinspace\ref{DM}(a) and (b)), the bilayer $\nu =1$ QH state collapses as $n_{\rm T}$ is increased.
This transition from the bilayer $\nu =1$ QH state to the compressible state is known to be induced by strong intralayer interactions for larger $n_{\rm T}$ \cite{Murphy}.
The QH state corresponds to the C phase because $B_{\parallel}$ is small.
When $B_\parallel $ is increased with $\theta $, a narrow QH region for the IC phase appears between the C and compressible states (Figs.\thinspace\ref{DM}(c) and (d)). 
We confirmed that this transition point exactly corresponds to the C-IC phase transition point by the activation energy measurements as a function of $\theta$ with fixed $n_{\rm T}$ (inset of Fig.\thinspace\ref{DM}(d))\cite{CICcomment5}.
%% \cite{SawadaPhysE, TerasawaPhysE, actiCIC}. 
As the sample is tilted further, the small $R_{xx}$ regions for the C and IC phases are separated  by the region having a relatively large $R_{xx}$ (Figs.\thinspace\ref{DM}(e)).
This appears as a sharp peak in the $R_{xx}$ plot along $\nu =1$ (Fig.\thinspace\ref{DM}(f)).

To investigate the origin of the $R_{xx}$ peak, we studied the anisotropic magnetotransport with respect to the angle $\phi$ between $B_{\parallel}$ and $I$ (Fig.\thinspace\ref{Anisotropy}(a)). Temperature was kept relatively high at $T=300$\,mK.
When $I$ is parallel to $B_{\parallel}$ ($\phi=0^{\rm o}$), the $R_{xx}$ shows a well-developed minimum for the bilayer $\nu =1$ QH state. However, as $\phi$ is increased, a peak in $R_{xx}$ grows and becomes maximum when $I$ is orthogonal to $B_{\parallel}$ ($\phi=90^{\rm o}$). Figure \ref{Anisotropy}(b) shows the magnetoresistance at the peak $R_{xx}^{\rm peak}$ as a function of $\phi$. The data is well fitted by a sinusoidal function: $R_{xx}^{\rm peak}=-A\cos 2\phi+B$. The amplitude $A=[R_{xx}(\phi=90^\circ )-R_{xx}(\phi=0^\circ )]/2$ is related to the anisotropic ratio $[R_{xx}(\phi=90^\circ )-R_{xx}(\phi=0^\circ )]/ R_{xx}(\phi=0^\circ )$. As $T$ is decreased, although $A$ decreases, $R_{xx}(\phi=0^\circ )$ decreases more rapidly and thus the anisotropic ratio increases [Fig.\,\ref{Anisotropy}(c)].

The observed $\phi $ dependence of $R_{xx}$ indicates the formation of a stripe-shaped structure parallel to $B_{\parallel}$ around the C-IC phase transition point.
We ascribe it to the domain structure of pseudospins, i.e. the soliton lattice.
When $\phi =0^\circ $, current flows parallel to pseudospin solitons.
As $\phi $ is increased to $90^\circ $, the number of solitons that cross the current increases.
The fact that $R_{xx}^{\rm peak}$ is minimum at $\phi =0^\circ $ and maximum at $\phi =90^\circ $ indicates that electrons are backscattered when they cross a pseudospin soliton.
This anisotropy is  reminiscent of the conventional spin-induced giant magnetoresistance (GMR)\cite{Baibich}, although in this system anisotropic spin-dependent scattering of the carriers is maximized when the magnetic moments of the ferromagnetic layers are anti-parallel.

\begin{figure}[t]
\includegraphics[width=\linewidth]{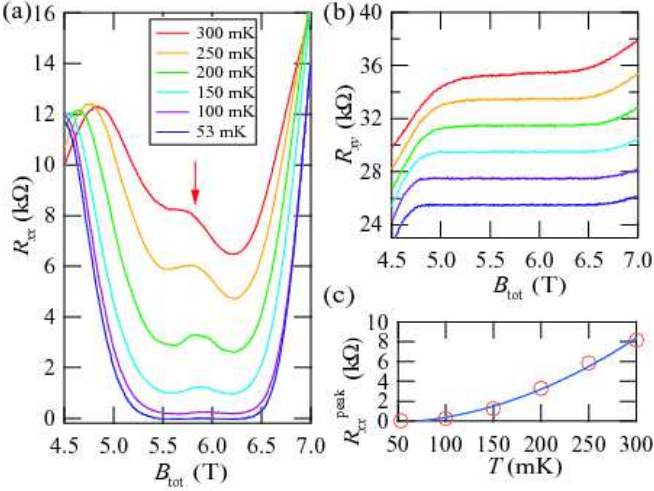}
\caption{(a) $R_{xx}$  and (b) $R_{xy}$  as a function of the total magnetic field for several temperatures at $\theta=50.3^{\rm o}$ and $\phi=90^{\rm o}$ for sample A.
The total electron density is $0.95 \times 10^{15}$\,${\rm m}^{-2}$. Hall resistance is offset by $2 {\rm k}\Omega$ for clarity.
(c) Temperature dependence of the magnetoresistance at the peak $R_{xx}^{\rm peak}$ indicated by the arrow in (a). }
\label{TDep}
\end{figure}

To get further insight into the dissipation mechanism of pseudospin solitons, we measured the temperature ($T$) dependence of $R_{xx}$ and $R_{xy}$. Figures \ref{TDep}(a) and (b) show $R_{xx}$ and $R_{xy}$ as a function of $B_{\rm tot}$, respectively, for several values of $T$. 
When $T=53$ mK, $R_{xx}$ is vanishingly small at $\nu=1$.  
As $T$ increases, a clear peak appears around $\nu =1$ for $T\ge 150$\,mK. 
The $R_{xx}^{\rm peak}$  is plotted in Fig.\thinspace\ref{TDep}(c) as a function of temperature.
On the other hand, the Hall plateau is well developed even at $T=300$\,mK  [Fig.\ref{TDep}(b)], showing that the anomalous peak in $R_{xx}$ occurs in the QH regime.

The temperature dependence of $R_{xx}$ at $\nu =1$ can be explained by fluctuations of the soliton lattice.
At the low temperature limit, solitons form a lattice, making a perfect periodic potential, that is, a Bloch state.
In such a system, there is no dissipation because there is no backscattering of electrons in a perfect periodic system.
This agrees with the experimental result of $R_{xx}\sim 0$ at low temperature.
At higher temperature, since the soliton lattice fluctuates thermally, the perfect periodicity of soliton lattice is lost and then backscattering occurs by individual fluctuating solitons.
As a result, $R_{xx}$ increases with temperature.
Fluctuations of solitons also affect the anisotropic ratio.
At higher temperature, not only the separation but also the orientation of solitons fluctuate, which make the anisotropy unclear. The anisotropic ratio decreases with increasing temperature and is consistent with the experimentally obtained temperature dependence of the anisotropic ratio (Fig.\ref{Anisotropy}(c)).
On the other hand,
fluctuations of solitons never affect the incompressibility of domains between solitons, and the Hall plateau develops even when backscattering occurs.

\begin{figure}[t]
\includegraphics[width=\linewidth]{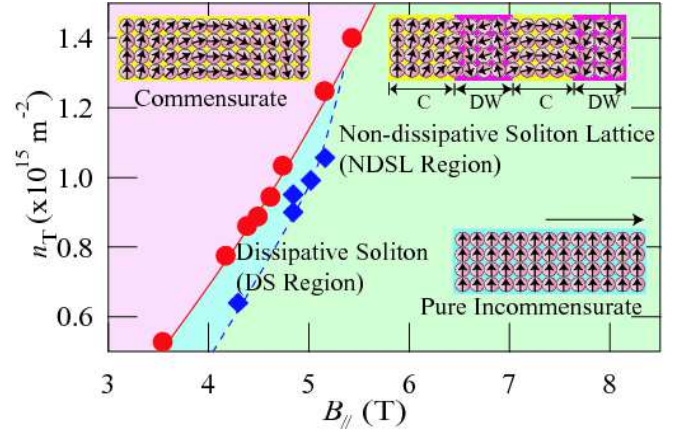}
\caption{Phase diagram of the bilayer $\nu=1$ QH state at $\sigma =0$ as a function of the in-plane magnetic field and the total electron density at $T=130$ mK. Circles with the solid line represent the upper boundary of C phase indicated by solid arrows in Fig.\thinspace\ref{DM}. Diamonds with the dashed line are the boundary between the non-dissipative soliton lattice (NDSL) region and the dissipative soliton (DS) region indicated by the dashed arrow in Fig.\thinspace\ref{DM}.
Data are for sample A. Schematic illustrations of pseudospins in each phase are also shown. (Arrows indicate the direction of pseudospins in the $xy$ plain.)}
\label{PD}
\end{figure}

We construct a phase diagram for the bilayer $\nu=1$ QH state in the $B_{\parallel}$ - $n_{\rm T}$ plane at a finite temperature $T=130$\,mK (Fig.\thinspace\ref{PD}).
The C-IC phase boundary (solid line) is obtained by collecting $(B_\parallel , n_{\rm T})$, where $R_{xx}$ starts to increase as indicated by solid arrows in Figs.\,\ref{DM}(d) and (f).
We divide the IC phase for $B_{\parallel} > B_{\parallel}^{\rm C}$ into two regions, that is, the non-dissipative and dissipative regions.
The boundary (dashed line) between the two regions is defined by $(B_\parallel , n_{\rm T})$ that gives a local minimum of the derivative of $R_{xx}$ with respect to $n_{\rm T}$ (dashed arrow in Fig.\thinspace\ref{DM}(f)).
The narrow dissipative region appears along the boundary to the C phase.

We discuss the $B_\parallel $ dependence of $R_{xx}$ in the IC phase.
At a finite temperature, $R_{xx}$ in the IC phase is determined by the stiffness of the soliton lattice, which is related to the density of soliton $n_{\rm S}$. Near the phase boundary to the C phase, $n_{\rm S}$ is small and the lattice is soft. Therefore, at a finite temperature, the fluctuations of solitons are large, leading to large $R_{xx}$.
As $B_\parallel $ is increased, the stiffness increases with $n_{\rm S}$ and, at some point, the lattice of soltions is formed.
Thus we refer to the dissipative and non-dissipative regions as a dissipative soliton  (DS) region and a non-dissipative soliton lattice (NDSL) region, respectively. The phase diagram shows that the DS region appears narrowly along the boundary to the C phase.
Theories \cite{Ezawa_b,Hanna} show that $n_{\rm S}$ starts to increase at $B_\parallel =B_\parallel ^{\rm C}$ and proliferates rapidly until the distance between solitons becomes comparable to the width of a soliton. The rapid increase in $n_{\rm S}$ explains the narrow DS region along the phase boundary.
Once the spacing between solitons are equal to the domain width, the interlayer phase difference of electrons is constant at every place, pseudospins are polarized and the system is regarded as a pure incommensurate phase, illustrated in Fig.\thinspace\ref{PD}. Since the $R_{xx}$ in the pure incommensurate phase also vanishes, it is observed experimentally in the NDSL region. 
Note that the Kosterlitz-Thouless (K-T) transition between the soliton lattice and liquid was theoretically predicted\cite{Hanna, ParkPRB}. According to the theory, it may be possible to translate the DS region into a soliton liquid phase.
However, our $R_{xx}$ data show no criticality. This may be due to the particularity of the K-T transition that any thermodynamic quantity does not jump at the transition point. More detailed theoretical work for the $n_{\rm S}$ and $T$ dependence of $R_{xx}$ would clarify the existence of the K-T transition.

\begin{figure}[t]
\includegraphics[width=\linewidth]{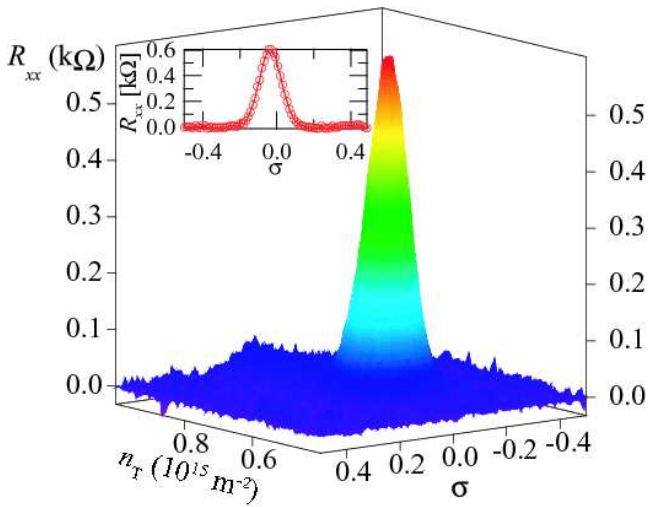}
\caption{(a) Surface plot of $R_{xx}$ as a function of the density
imbalance parameter 
$\sigma$
 and the total density $n_{\mathrm{T}}$ for sample A. $R_{xx}$ was measured by sweeping both $\sigma$ and $n_{\rm T}$ while keeping the filling factor $\nu=1$ at $\theta=57.9^\circ$ and $\phi=90^{\circ}$. Temperature is 100 mK. Inset:  $R_{xx}$ as a function of $\sigma$ at $n_{\mathrm{T}}=0.61\times10^{15}$\,${\rm m}^{-2}$.}
\label{SMF}
\end{figure}

Finally, we investigate effects of the density imbalance $\sigma $ between the two layers. Figure\thinspace\ref{SMF} shows $R_{xx}$ in a surface plot as a function of $\sigma$ and $n_{\mathrm{T}}$. The data were taken by sweeping both $\sigma$ and $n_{\rm T}$ while keeping the filling factor $\nu=1$ at $\theta=57.9^\circ$ and $\phi=90^{\circ}$. Pseudospin solitons appear as a peak in $R_{xx}$ around $\sigma =0$ and $n_{\mathrm{T}}=0.61\times10^{15}$\,${\rm m}^{-2}$. The inset of Fig.\thinspace\ref{SMF} shows the slice at $n_{\mathrm{T}}=0.61\times10^{15}$\,${\rm m}^{-2}$.  For $|\sigma |>0.2$, $R_{xx}$ almost vanishes, which indicates that pseudospin solitons are unstable.
The instability would be due to the reduction of the pseudospin stiffness proportional to $1-\sigma ^2$ \cite{Cote_SL,Hanna}. 
However, this dependence is not sufficient to explain the observed strong $\sigma $ dependence of $R_{xx}$. 
Further quantitative theoretical investigation is needed to reveal the structure of pseudospin solitons in an off-balanced system.

In conclusion, we have carried out magnetotransport measurements around the C-IC transition in the bilayer $\nu =1$ QH state. We found an anomalous peak in $R_{xx}$ near the C-IC transition point. This peak has a highly anisotropic nature, which indicates that pseudospin solitons are formed along the direction of $B_{\parallel}$. 
The temperature dependence of $R_{xx}$ and $R_{xy}$ reveals that thermally fluctuated pseudospin solitons  solidify into a rigid soliton lattice at low temperature. 
The phase diagram constructed experimentally in the $n_{\rm T}$-$B_{\parallel}$ plane shows that the dissipation occurs only in a narrow region just near the C phase. We also found that pseudospin solitons are very sensitive to the charge imbalance.

We are grateful to T. Saku for growing the heterostructures, and C. B.
Hanna, T. Sekikawa, Y. Ogasawara, T. Satoh, Y. Maeno and K. Muraki for fruitful discussions. This research was supported in part by Grants-in-Aid for the Scientific Research (Nos. 14010839, 18740181, 18340088, and 18043012).
%%and a 21st Century COE Program Grant of the International COE of Exploring New Science Bridging Particle-Matter Hierarchy from the Ministry of Education, Culture, Sports, Science and Technology of Japan.  
Authors D.T. and K.I. are grateful for the Research Fellowships of the Japan Society for the Promotion of Science for Young Scientists.

\bibliographystyle{unsrt}
%\bibliographystyle{plain}
%\bibliography{AFbibrr}

\begin{thebibliography}{99}
\bibitem{noteDSAS}
We obtained qualitatively the same results from the two samples with slightly
  different $\Delta_{\rm SAS}$. Quantitatively, overall features occur at
  higher $\theta$ in the sample with larger $\Delta_{\rm SAS}$.
\bibitem{CICcomment5}
A rotation of the sample in the mixing chamber increases the temperature,
  preventing the detailed experiment around the C-IC transition by changing
  $\theta$ with fixed $n_{\rm T}$. Accordingly, most of the data are taken by
  changing $n_{\rm T}$ and $B_{\rm tot}$ with fixed $\theta$.
\end{thebibliography}
% Produces the bibliography via BibTeX.

\hyphenation{Post-Script Sprin-ger}

\end{document}